\documentclass[preprint,12pt,authornum]{elsarticle}
\usepackage{multirow}
\usepackage{amsmath}
\usepackage{amssymb}
\usepackage{graphicx}
\usepackage{subcaption}
\usepackage{caption}
\usepackage[utf8]{inputenc}
\usepackage{geometry}
\usepackage{xcolor}
\usepackage{enumerate}
\usepackage{algorithm}
\usepackage{algorithmic}
\usepackage{setspace}
 \usepackage{url}
\usepackage[colorlinks = true,
            linkcolor = blue,
            urlcolor  = blue,
            citecolor = blue,
            anchorcolor = blue]{hyperref}
\usepackage{natbib}
\journal{EESD}

\begin{document}

\begin{frontmatter}



\title{The Importance of Corner Frequency in Site-Based Stochastic Ground Motion Models}
% On the importance of corner frequency 

\author[inst1]{Maijia Su\corref{cor2}}
\author[inst2]{Mayssa Dabaghi}
\author[inst1]{Marco Broccardo\corref{cor1}}

\affiliation[inst1]{organization={Department of Civil, Environmental and Mechanical Engineering, University of Trento, Trento},%Department and Organization
            country={Italy}}
\affiliation[inst2]{organization={Department of Civil and Environmental Engineering,  American University of Beirut, Beirut},%Department and Organization
            country={Lebanon}}

\cortext[cor1]{Correspondence: marco.broccardo@unitn.it}
\cortext[cor2]{Correspondence: maijia.su@unitn.it}
\begin{abstract}
Synthetic ground motions (GMs) play a fundamental role in both deterministic and probabilistic seismic engineering assessments.
This paper shows that the family of filtered and modulated white noise stochastic GM models overlooks a key parameter---the high-pass filter's corner frequency, \(f_c\). 
In the simulated motions, this causes significant distortions in the long-period range of the linear-response spectra and in the linear-response spectral correlations. 
To address this, we incorporate \(f_c\) as an explicitly fitted parameter in a site-based stochastic model. We optimize \(f_c\) by individually matching the long-period linear-response spectrum (i.e., $Sa(T)$ for $T \geq 1$s) of synthetic GMs with that of each recorded GM. We show that by fitting \(f_c\) the resulting stochastically simulated GMs can precisely capture the spectral amplitudes, variability (i.e., variances of  $\log(Sa(T))$), and the correlation structure (i.e., correlation of $\log(Sa(T))$ between distinct periods $T_1$ and $T_2$) of recorded GMs. To quantify the impact of $f_c$, a sensitivity analysis is conducted through linear regression. This regression relates the logarithmic linear-response spectrum ($\log(Sa(T))$) to  seven GM parameters, including the optimized \(f_c\).  The results indicate that the variance of \(f_c\) observed in natural GMs, along with its correlation with the other GM parameters, accounts for 26\% of the spectral variability in long periods. Neglecting either the \(f_c\) variance or \(f_c\) correlation typically results in an important overestimation of the linear-response spectral correlation.

\end{abstract}




\end{frontmatter}

% \linenumbers

\section{Introduction}
\noindent 
A proper selection or generation of ground motion (GM) time series is vital for designing earthquake-resistant buildings and conducting seismic risk assessments. 
Efforts to synthesize GMs include site-based stochastic models, which have evolved since the 1950s, progressing from stationary white-noise models to fully temporal- and spectral-nonstationary models. 
It is critical that synthetic GMs capture the essential characteristics and natural variability of recorded GMs. In recent years, site-based stochastic GMs \cite{rezaeian_simulation_2012,dabaghi_simulation_2018,vlachos_predictive_2018} have gained widespread attention  due to their computational efficiency, straightforward formulations, and reliance on input that consists of readily available information about earthquake source, path, and site characteristics (e.g., the earthquake magnitude, source-to-site distance, and shear-wave velocity of the site).

Site-based GM models are directly fitted to a recorded GM catalog of interest, which includes GMs selected from a specified range of earthquake scenarios defined by their source, path and site characteristics. Specifically, GMs in the catalog are individually fitted using a prescribed parametric GM model. This study focuses on stochastic GM models based on simulating modulated and filtered white noises. The random process of the filtered white-noise can be simulated in two ways: in the time domain by solving a linear filter excited by Gaussian white noise \cite{rezaeian_stochastic_2008}  or in the frequency domain by using spectral representations based on Priestley’s evolutionary theory \cite{broccardo2017spectral}. Typically, these filtered processes are subsequently refined using a high-pass filter, ensuring (1) that artificially generated GMs achieve zero residual velocity and displacement at the end of the process and, more importantly, (2) that their low-frequency contents are adjusted to match those of recorded GMs.

Previous models have treated the corner frequency, $f_c$, of the high-pass filter as a constrained parameter and have used various methods to estimate its value, including using a constant value of 0.1Hz \citep{conte_fully_1997,rezaeian_simulation_2012} or 0.2Hz \citep{vlachos_predictive_2018}, values from the flat-file of the PEER-NGA database \citep{rezaeian_stochastic_2008}, and values predicted using equations based on earthquake magnitude \citep{dabaghi_simulation_2018,sabetta_simulation_2021} or earthquake motion duration \citep{liao_physically_2006,vlachos_multi-modal_2016}. 
This results in synthetic GMs with distorted long-period linear-response spectra and incorrect linear-response spectral correlation structures, as illustrated in Section \ref{Section2}. In this context, we show that $f_c$ plays a key role and needs to be a free parameter of the parametric stochastic GM model. Consequently, Section \ref{Section3} introduces a novel fitting procedure to optimize \(f_c\). Specifically, the procedure requires matching the long-period content of the simulated linear-response spectra with those computed from recorded motions. The resulting synthetic GMs with optimized \(f_c\)  also have their linear-response spectral correlation structure consistent with the recorded motions.
In Section \ref{Section4}, a comprehensive sensitivity analysis is conducted through linear regression to gain fundamental insights into the impact of the corner frequency on linear-response spectra and spectral correlation structures.

\section{Effect of $f_c$ on the Linear-Response Spectra of Simulated Motions}
\label{Section2}
\noindent
This paper builds upon the findings of a previous study by Broccardo and Dabaghi \cite{broccardo2019preliminary}, which compared statistical quantities of  the 5\%-damped elastic pseudo-acceleration response spectrum ($S_a(T)$) between a recorded GM catalog\footnote{The recorded catalog has 71 GMs which are downloaded from the PEER NGA-West2 database, see details in \cite{rezaeian_simulation_2012,broccardo2019preliminary}. The search conditions use reverse fault type, source-to-site distances $10 \sim 90$ km, magnitudes $6 \sim 7.6$, and soil shear-wave velocity $V_{S30}\geq600$ m/s.} and its corresponding synthetic catalogs. The study found that the synthetic catalog generated using the temporal-based GM model proposed by Rezaeian and Der Kiureghian \cite{rezaeian_simulation_2012} overestimates  spectral correlations (i.e., Pearson's correlation coefficients of $\log(S_a(T))$ at periods $T_1$ and $T_2$). In contrast, the frequency-domain equivalent model \cite{broccardo2017spectral} of the GM model \cite{rezaeian_simulation_2012} exhibits improved correlations. This paper reveals that the choice of GM simulation method in the time or frequency domain is not critical. The improved correlation pattern arises from a different approach in estimating  $f_c$ in \cite{broccardo2017spectral}.  Moreover, the improper selection of \(f_c\) not only skews the spectral correlation but also affects the spectral marginal (i.e., the marginal probability distribution of \(S_a(T)\)) at periods   $T \geq 1$s.

\begin{figure}[htb]
\centering
\begin{subfigure}[hb]{0.95\textwidth}
   \includegraphics[width=1\linewidth]{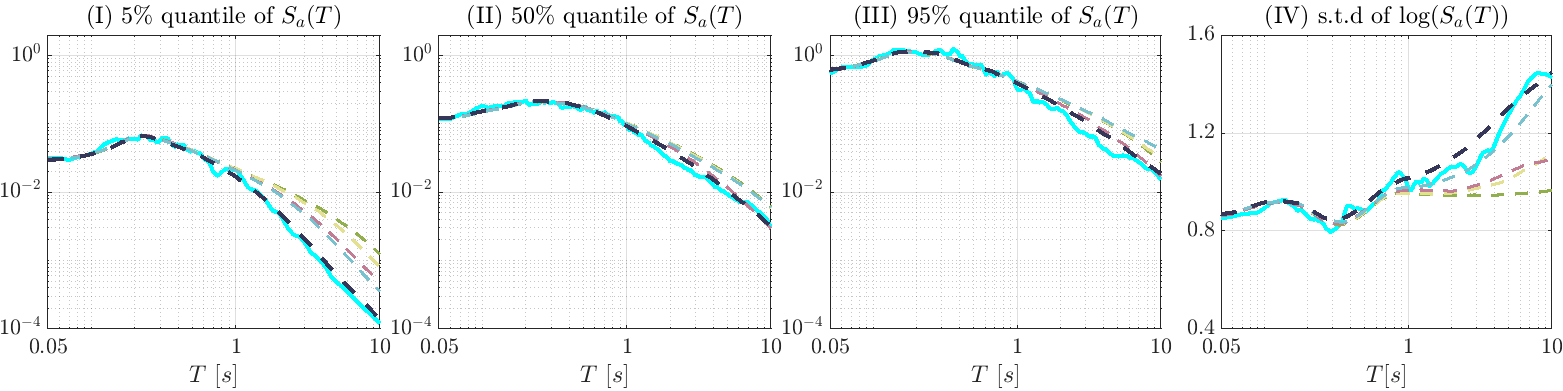}
   \caption{Temporal-based model}
   \label{fig1a} 
\end{subfigure}

\begin{subfigure}[hb]{0.95\textwidth}
   \includegraphics[width=1\linewidth]{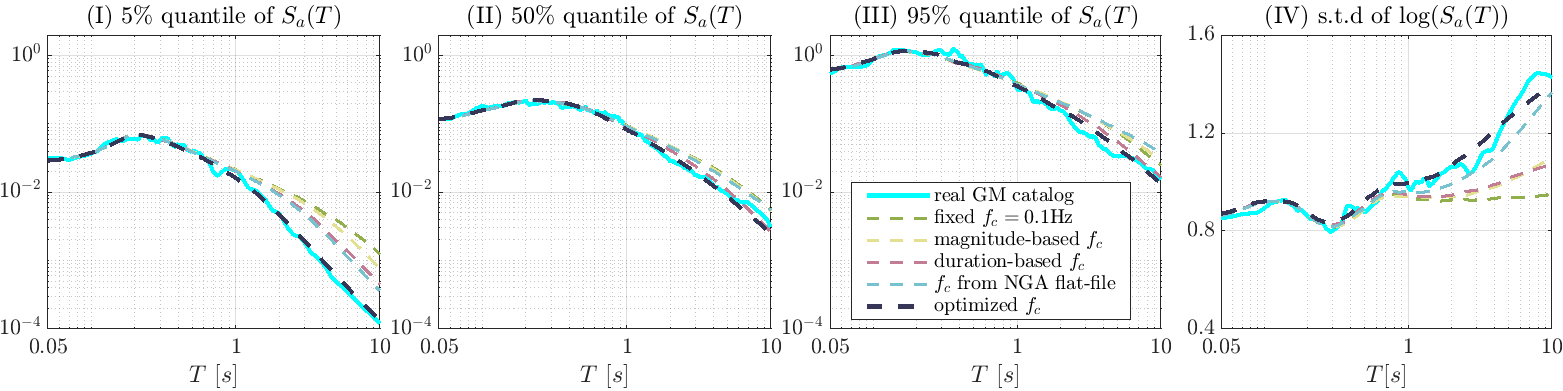}
   \caption{Spectral-based model}
   \label{fig1b}
\end{subfigure}
\caption{Comparing statistics ( 5\% quantile, 50\% quantile, 95\% quantile and standard deviation) of 5\%-damped elastic response spectra between a real GM catalog and 5 corresponding synthetic catalogs using different $f_c$ estimation methods: (a) Synthetic catalogs simulated using the temporal-based model \cite{rezaeian_simulation_2012}; (b) Synthetic catalogs simulated using the spectral-based modal \cite{broccardo2017spectral}}
\label{fig1} 
\end{figure}

\begin{figure}[htb]
\centering
\begin{subfigure}[b]{0.95\textwidth}
   \includegraphics[width=1\linewidth]{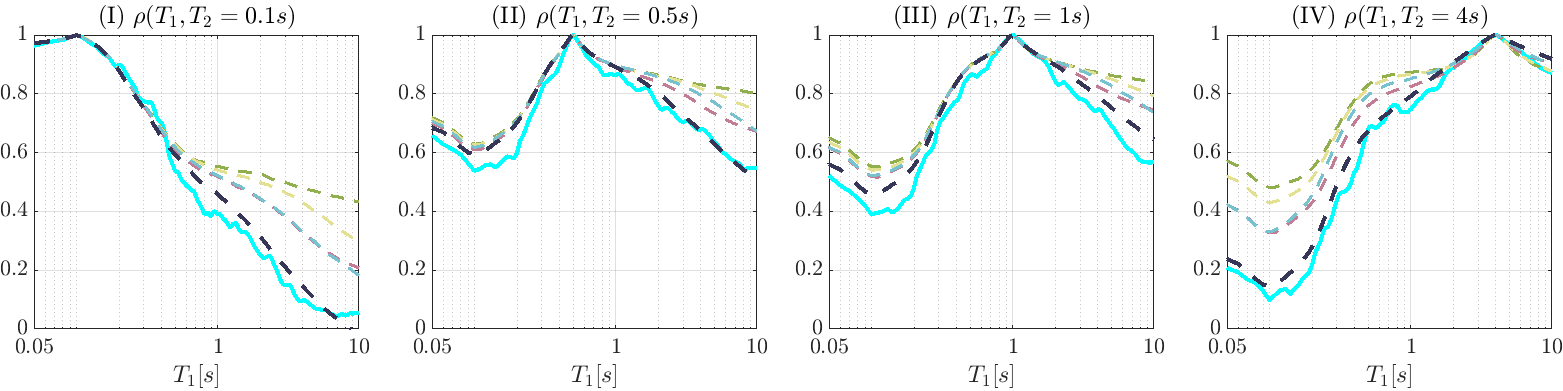}
   \caption{Temporal-based model}
   \label{fig2a} 
\end{subfigure}
\begin{subfigure}[b]{0.95\textwidth}
   \includegraphics[width=1\linewidth]{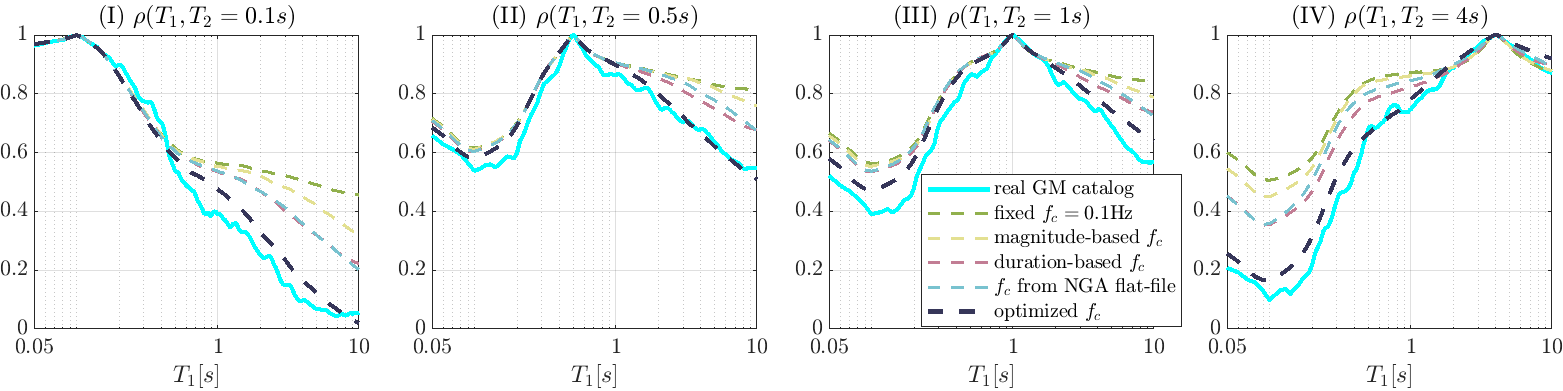}
   \caption{Spectral-based model}
   \label{fig2b}
\end{subfigure}
\caption{ Comparing the spectral correlations between a real GM catalog and 5 corresponding synthetic catalogs using different $f_c$ estimation methods: (a) Synthetic catalogs simulated using the temporal-based model \cite{rezaeian_simulation_2012}; (b) Synthetic catalogs simulated using the spectral-based modal \cite{broccardo2017spectral} }
\label{fig2} 
\end{figure}

% To illustrate our findings, we compare the response spectrum statistics of synthetic GM catalogs obtained using both the temporal- and spectral-based model formulations, along with various methods of estimating $f_c$, to those of the corresponding recorded GM catalog.
To illustrate our findings, we compare the $S_a(T)$ statistics of synthetic GM catalogs to those of the corresponding recorded GM catalog studied in \cite{rezaeian_simulation_2012,broccardo2019preliminary}. These synthetic catalogs are obtained using both the temporal- and spectral-based model formulations \cite{rezaeian_simulation_2012,broccardo2017spectral}, along with various methods of estimating $f_c$. Figure \ref{fig1}  and Figure \ref{fig2} respectively summarize the improved spectral marginal and correlation when using the optimized $f_c$ method presented in Section \ref{Section3}  in comparison to four other traditional $f_c$  methods.  These traditional methods involve using a constant value $f_c=0.1$Hz \cite{rezaeian_simulation_2012},  an earthquake-magnitude-based prediction equation \cite{dabaghi_simulation_2018}, a GM-duration-based prediction equation \cite{liao_physically_2006} and the values provided in NGA flat-file \cite{rezaeian_stochastic_2008}. The comparisons show no significant difference between using the temporal-based and the spectral-based GM model. The differences arise from the choice of the $f_c$ computation method. Specifically, Figure \ref{fig1} compares four key features of the spectral marginal: the 5\% quantile, 50\% quantile (i.e., median), 95\% quantile, and standard deviation (s.t.d) of $\log(S_a(T))$. The comparison reveals that all \(f_c\) estimation methods accurately capture the response spectrum at short periods (\(T \leq 1\) s); however, only the optimized method consistently provides a good fit at long periods (\(T > 1\) s). Traditional \(f_c\) estimation methods tend to overestimate the quantiles of the spectral responses but underestimate the variance of the spectral marginal. Figure \ref{fig2} compares the spectral correlations between periods $T_1 \in [0.05,10]$s and four fixed periods $T_2 = 0.1,0.5,1,4$s. Overall, the optimization-based method exhibits the closest fit, while the fixed $f_c$ results in large over-estimations of the correlation between short- and long-period ranges.

%% Figure 2

\section{Optimized Corner Frequency Fitting}
\label{Section3}
\noindent
This section proposes a method to optimize the fitting of parameter $f_c$ for stochastic GM models (e.g., \cite{rezaeian_simulation_2012,broccardo2017spectral,vlachos_predictive_2018} ) that require a high-pass filter to eliminate the unrealistic low-frequency contents. To make this paper self-contained, this section  briefly introduces the temporal-based \cite{rezaeian_stochastic_2008} and spectral-based \cite{broccardo2017spectral} stochastic GM models; readers can refer to their original papers for more details.

We integrate the temporal-based \cite{rezaeian_stochastic_2008} and spectral-based \cite{broccardo2017spectral} stochastic GM models into a consistent framework. Both models start by applying a time-varying linear filter to a Gaussian white-noise excitation. The resulting colored noise has spectral nonstationarity due to the time-varying filter. Next, the signal is adjusted with a time-modulating function to achieve temporal nonstationarity. Finally, a high-pass filter is applied to adjust the low-frequency contents of the simulations to match a specific recorded GM and assure zero velocity and displacement at the end of the motion. The distinction between the two models lies in the definition of the time-varying linear filter:  temporal-based models utilize an evolutionary impulse response function $h_{X}(t-\tau|\boldsymbol{\theta}_F(\tau))$, while spectral-based models use an evolutionary frequency-response function $H_{XX}(\omega|\boldsymbol{\theta}_F(t))$. The parameter vector $\boldsymbol{\theta}_F(t)$ determines the transient properties of the linear filter. In summary, both stochastic GM models follow a four-step procedure, including: 
\begin{itemize}
      \item \textit{Step 1:} Simulate a filtered white noise process $X_1(t)$
\begin{itemize}
    \item either in the \textit{time domain}:   $X_1(t) = \int_0^{t} h_{X}(t-\tau|\boldsymbol{\theta}_F(\tau))d W(\tau)$,
    \item or in the \textit{frequency domain}:   $X_1(t) = \int_{-\infty}^{+\infty}H_{XX}(\omega|\boldsymbol{\theta}_F(t))\exp(-i\omega t)d\hat W(\omega),  $\\
where $W(\tau)$ is a real Gaussian white-noise process, and $\hat W(\omega)$ is  a zero-mean complex valued process defined as the Fourier spectrum of $W(\tau)$.
\end{itemize}
    \item \textit{Step 2:} Normalize $X_1(t)$ to a unit-variance process, denoted as $X_2(t)=X_1(t)/\sigma_{X_1}(t)$, where $\sigma_{X_1}(t)$ is the process variance of $X_1(t)$.
    \item \textit{Step 3:} Envelop $X_2(t)$ with a time-modulating function $q(t|\boldsymbol{\theta}_T)$ and obtain $X_3(t) = q(t|\boldsymbol{\theta}_T)X_2(t)$, where $\boldsymbol{\theta}_T$ is the vector of parameters that control the temporal energy distribution.
    \item \textit{Step 4:} Convolve ${X}_3(t)$ with a high-pass filter $h_f(t|f_c)$ and the filter outputs the simulated GM.
\end{itemize}

The above 4-step procedure includes three parametric models: $h_{X}(t-\tau|\boldsymbol{\theta}_F(\tau))$ (or $H_{XX}(\omega|\boldsymbol{\theta}_F(t))$), $q(t|\boldsymbol{\theta}_T)$, and $h_f(t|f_c)$. This study chooses $h_f(t|f_c)$ as the impulse response function of a critically damped oscillator, represented by \(h_f(t|f_c) = t\exp(-2{\pi}f_c)\). The parameter estimations for the first two models are thoroughly studied in \cite{rezaeian_stochastic_2008,broccardo2017spectral} and they are assumed to be already fitted in this study. Next, we focus on fitting the final parameter, the corner frequency $f_c$, by proposing an optimization criterion. 

\begin{figure}[!htb]
   \begin{minipage}{0.48\textwidth}
     \centering
     \includegraphics[width=\textwidth]{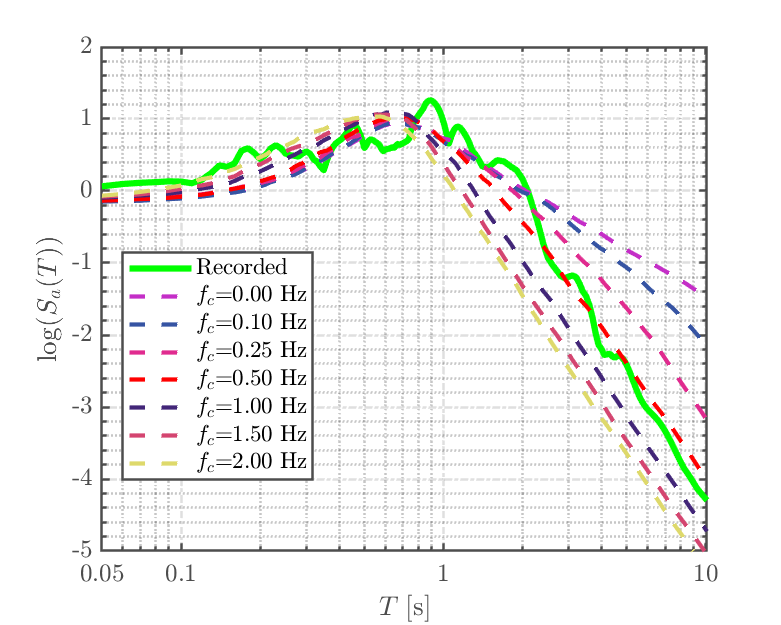}
     \caption{Comparison of $\log(Sa(T))$ of a recorded GM and \(\mathbb{E}(\log(S_a(T)))\) of synthetic GMs when setting the corner frequency $f_c$ to different values.}
     \label{fig_sec2_1}
 \end{minipage}\hfill
   \begin{minipage}{0.48\textwidth}
     \centering
     \includegraphics[width=\textwidth]{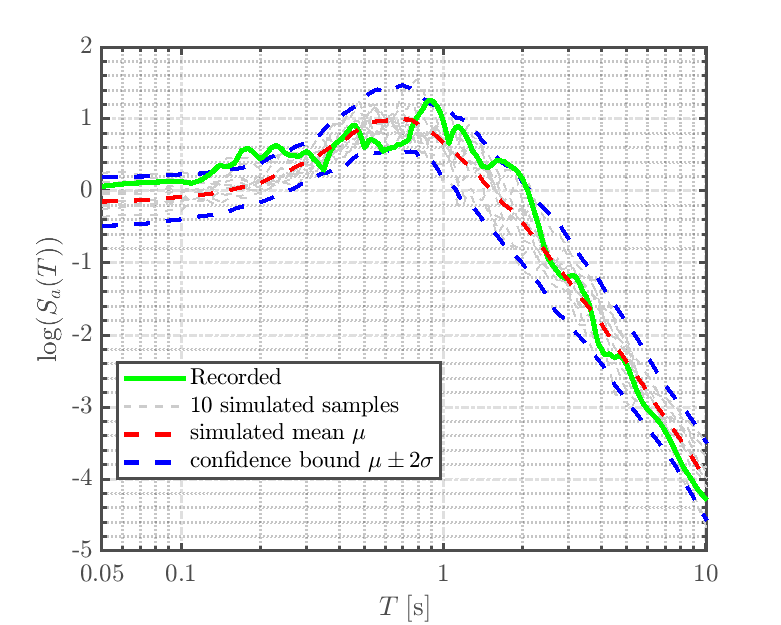}
     \caption{Comparison of the logarithmic spectrum response $\log(Sa(T))$ of a recorded GM and synthetic GMs generated from the fitted stochastic GM model using $f_c=0.5$Hz.}
     \label{fig_sec2_2}
   \end{minipage}
\end{figure}

The impact of applying a high-pass filter is well-known: it slightly alters the simulated linear-response spectrum at short periods (\(T<1\)s) but significantly affects long-period responses. For example, Figure \ref{fig_sec2_1} presents the spectrum \(S_a(T)\) of a GM recording from the Chi-Chi, Taiwan earthquake (NGA\#1517 downloaded from the PEER-west2 database). The figure also shows the mean logarithmic spectrum \(\mathbb{E}(\log(S_a(T)))\) of the corresponding simulated ground motions, generated using fitted model parameters $\hat{\boldsymbol{\theta}}_F(t)$ and $\hat{\boldsymbol{\theta}}_T$, with various \(f_c\) values ranging from \(0\) to \(2\) Hz\footnote{When $f_c=0$Hz, the high-pass filter is not used.}. In Figure \ref{fig_sec2_1}, it is apparent that setting \(f_c=0.5\) Hz provides the closest match to the recorded spectrum. Further detailed results with \(f_c=0.5\) Hz are shown in Figure \ref{fig_sec2_2}. The figure illustrates 10 samples from the fitted GM model together with the mean and $\pm$2$\sigma$-confidence bounds of 100 GM samples. These observations in Figure \ref{fig_sec2_1} and \ref{fig_sec2_2} inspire us to optimize $f_c$  such that the recorded spectrum best falls within the confidence bounds of the simulated spectra.

The mismatch between the recorded and simulated spectra is quantified as their relative bias accumulated over the long-period region (i..e,   $T\in[1,10]$).  Let  $S_a^{\mathrm{real}}(T)$ and $S_a^{\mathrm{sim}}(T|f_c)$ denote the as-recorded and corresponding simulated spectra, respectively. The simulated spectrum  $S_a^{\mathrm{sim}}(T|f_c)$ is conditional on $f_c$ and is a random variable due to the white-noise excitation. The optimization objective  is to minimize the summation of the relative bias, denoted as:
\begin{equation}
\label{eq_sec2_1}
    \epsilon(f_c) = \left| \int_{1}^{10}   \frac{  \log(S_a^\mathrm{real}(T)) - \mu[\log(S_a^\mathrm{sim}(T|f_c))]}{\sigma[\log(S_a^\mathrm{sim}(T|f_c))]}   d\log(T) \right|,
\end{equation}
where $\mu[\cdot]$  and  $\sigma[\cdot]$ are operators that respectively yield the mean and standard deviation of the input random variables. The bias is measured on the logarithmic scale and normalized by the standard deviation.  In practice, we solve the logarithmic integrals in Eq.\eqref{eq_sec2_1}  by summing the biases on 30 logarithmically spaced points within the integral region. Additionally, the statistical quantities in  Eq.\eqref{eq_sec2_1}  are estimated through Monte Carlo Simulation by drawing 100 samples from the fitted GM model. Finally, we determine the optimized value of $f_c$ that minimizes Eq.\eqref{eq_sec2_1} within a grid ranging from $[0,2]$ Hz, with an interval of $\Delta f_c = 0.01$ Hz. For the recorded GM used in Figure \ref{fig_sec2_1}, we obtain $f_c= 0.54$ Hz. 

It is important to note that the proposed optimized \(f_c\) fitting method is versatile and can be applied to other high-pass filters, such as the Butterworth filter. To adapt it for different filters, adjustments can be made to the grid range and the discretization step size \(\Delta f_c\).

\section{Sensitivity of Response Spectrum to $f_c$}
\label{Section4}
\noindent
In this section, we conduct a sensitivity analysis to quantify the impact of \(f_c\)  variability on the variability of the linear-response spectrum. This analysis utilizes a linear regression model, where seven GM parameters are used as input variables and $\log(Sa(T))$ as the output. Among these parameters, one is \(f_c\), and the remaining six are parameters from the stochastic GM model proposed by Rezaeian and Der Kiureghian \cite{rezaeian_simulation_2012}. These six GM parameters include the logarithmic Arias intensity (\(\log(AI)\)), effective duration (\(D_{5-95}\)), time taken to reach the middle of the strong-shaking phase (\(t_{mid}\)), and three key GM parameters at \(t_{mid}\): the predominant frequency $\omega(t_{mid})$, the rate of change of the filter frequency \(\omega'(t_{mid})\), and the filter bandwidth \(\zeta(t_{mid})\).  

The linear regression model is represented as follows:
\begin{equation}
    \label{eq_sec3_1}
    Y = \beta_0 + \sum\limits_{n=1}^{7}\beta_n \theta_n + \epsilon = \beta_0 +\boldsymbol{\beta}^\mathrm{T}\boldsymbol{\theta} + \epsilon,
\end{equation}
where the input random vector refers to $\boldsymbol{\theta}= [ \log(AI),D_{5-95},t_{mid},\omega(t_{mid}),\omega'(t_{mid}),\zeta(t_{mid}),f_c   ]^{\mathrm{T}}$. The input vector covariance matrix is  $\Sigma_{\boldsymbol{\theta\theta}}$, with the $n$-th diagonal entry representing the variance  $\sigma^2_{\theta_n}$. The regression error $\epsilon$ is a zero-mean Gaussian random variable, incorporating two error sources: truncation error due to linearization and unaccounted GM information excluded from the parameters $\boldsymbol{\theta}$. The regression is repeated for multiple spectral ordinates by setting $Y = \log(Sa(T))$ for $T \in [0.05,10]$s. The linear regression in Eq.\eqref{eq_sec3_1} facilitates the decomposition of the spectral variance $\operatorname{Var}(Y(T))$ into two terms and the spectral covariance $\operatorname{Cov}(Y(T_1),Y(T_2))$ into four terms:
\begin{equation}
    \label{eq_sec3_2}
    \operatorname{Var}(Y(T)) = \boldsymbol{\beta}^{\mathrm{T}}(T)\Sigma_{\boldsymbol{\theta\theta}}\boldsymbol{\beta}(T) + \operatorname{Var}(\epsilon(T)),
\end{equation}
\begin{equation}
    \label{eq_sec3_3}
    \begin{aligned}
          \operatorname{Cov}(Y(T_1),Y(T_2)) & = \boldsymbol{\beta}^{\mathrm{T}}(T_1)\Sigma_{\boldsymbol{\theta\theta}}\boldsymbol{\beta}(T_2) + \boldsymbol{\beta}^{\mathrm{T}}(T_1)\operatorname{Cov}(\boldsymbol{\theta},\epsilon(T_2))  \\
          &+ \boldsymbol{\beta}^{\mathrm{T}}(T_2)\operatorname{Cov}(\boldsymbol{\theta},\epsilon(T_1)) + \operatorname{Cov}(\epsilon(T_1),\epsilon(T_2))
    \end{aligned}.
\end{equation}

In this study, the regression parameters $\boldsymbol{\beta}$  are fitted by the ordinary least-squares method.  The variance $\operatorname{Var}(\epsilon(T))$ and covariances ($\Sigma_{\boldsymbol{\theta\theta}}$, $\operatorname{Cov}(\epsilon(T_1),\epsilon(T_2))$ and  $\operatorname{Cov}(\boldsymbol{\theta},\epsilon(T))$) are computed from the GM parameters and regression errors using the corresponding empirical (co)variance formula. 

We illustrate this analysis for the same GM catalog as used in Figure \ref{fig1} and \ref{fig2}. The GM parameters $\log(AI)$, $D_{5-95}$  and $t_{mid}$ can be directly computed by their definitions. The parameters $\omega(t_{mid}),\omega'(t_{mid}),\zeta(t_{mid})$ are extracted from the short-time Thomson's multiple-window spectrum of GMs (see details in \cite{conte_fully_1997,broccardo2017spectral}).  The final parameter $f_c$ is optimized using the method proposed in Section \ref{Section3}. The statistical observations of these extracted parameters, along with their corresponding fitted probabilistic models, are provided in \ref{AppendixA}. 

\begin{figure}[!htb]
   \begin{minipage}{0.48\textwidth}
     \centering
     \includegraphics[width=1\linewidth]{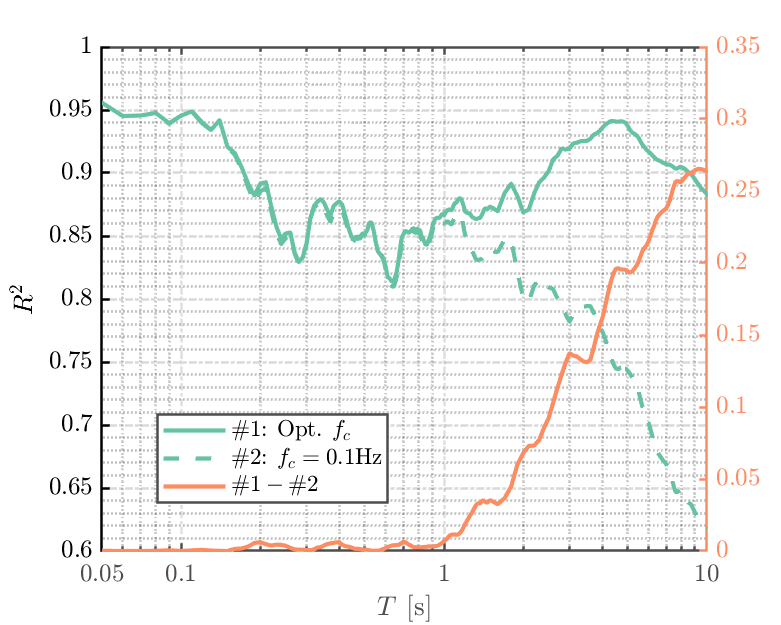}
     \caption{Coefficient of determination $R^2$}
     \label{fig_sec3_1}
 \end{minipage}\hfill
   \begin{minipage}{0.48\textwidth}
     \centering
     \includegraphics[width=1\linewidth]{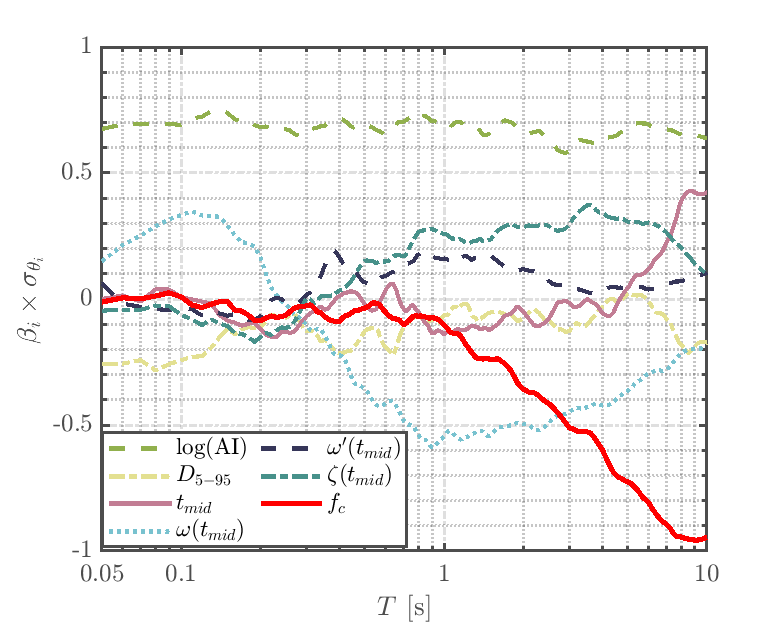}
     \caption{Weighted regression coefficients ${\beta}_n \times \sigma_{\theta_n}$}
     \label{fig_sec3_2}
   \end{minipage}
\end{figure}

The fitting outcomes over the spectrum ordinate \(T\in[0.05,10]\)s are depicted in Figure \ref{fig_sec3_1} and \ref{fig_sec3_2}. Figure \ref{fig_sec3_1} illustrates the coefficient of determination (\(R^2\))  and compares two cases of $f_c$: the optimized values versus a constant value of 0.1Hz. A constant $f_c$ implies neglecting the $f_c$ variance (i.e., $\sigma^2_{\theta_7}=0$) and its covariance with the other six parameters, i.e., $[{\Sigma}_{\boldsymbol{\theta\theta}}]_{7,n}=[{\Sigma}_{\boldsymbol{\theta\theta}}]_{n,7}=0$ for $n \in [1,2,...,6]$. The coefficient \(R^2\) serves as an evaluation metric for the linear models, indicating the proportion of the output variance that the linear regression can predict, i.e., \(R^2  = \boldsymbol{\beta}^{\mathrm{T}}\Sigma_{\boldsymbol{\theta\theta}}\boldsymbol{\beta} / \operatorname{Var}(Y)\).  The figure indicates a significant enhancement in accounting for the total variance for periods \(T \geq 1\)s when using the optimized \(f_c\). This improvement becomes more substantial with the increase of \(T\), reaching up to 26\% at \(T=10\)s. Moreover, by using the optimized $f_c$, a linear relationship among the input-output pairs already explains \(81\% \sim 95\%\) of spectral variances; the remaining variance due to the regression error is small\footnote{\ref{AppendixB} additionally shows that the linear-regression covariance $\boldsymbol{\beta}^{\mathrm{T}}(T_1)\Sigma_{\boldsymbol{\theta\theta}}\boldsymbol{\beta}(T_2)$ in Eq.\eqref{eq_sec3_3} explains \(70\% \sim 130\%\) of the total covariance $ \operatorname{Cov}(Y(T_1),Y(T_2))$.}.  
Figure \ref{fig_sec3_2} compares the weighted regression coefficients  \({\beta}_n \times \sigma_{\theta_n}\)  of the seven GM parameters. The sum of squares of these coefficients equals the regression variances  (i.e., $\boldsymbol{\beta}^{\mathrm{T}}\Sigma_{\boldsymbol{\theta\theta}}\boldsymbol{\beta}$) when the input variables are uncorrelated. The figure again indicates that the coefficient related to \(f_c\) becomes pronounced after \(T=1\)s and gains more significance with increasing \(T\). At long period $T>5$s, the $f_c$ coefficient outweighs the coefficients of the other 6 GM parameters.

% \begin{figure}[htb]
% \begin{minipage}[b]{0.48\textwidth}
%    \centering
%    \label{fig_sec3_1} 
%    \includegraphics[width=1\linewidth]{Fig_sec3_1.png}
%    \caption{coefficient of determination $R^2$}
% \end{minipage} 
% \begin{minipage}[b]{0.48\textwidth}
%     \centering
%    \label{fig_sec3_2}
%    \includegraphics[width=1\linewidth]{Fig_sec3_3.png}
%    \caption{Standard regression coefficients ${\beta}_n \times \sigma_{\theta_n}$}
% \end{minipage}
% \end{figure}

% Next, we investigate  the  fc variance and fc covariance 
% In Figure \ref{fig1} and \ref{fig2},  we find that the impact of fc on the spectrum in the GM models and the linear regression model 
% by respectively setting  $f_c=0.1$Hz   (i.e., by setting \([{\Sigma}_{\boldsymbol{\theta\theta}}]_{1:7,7}=[{\Sigma}_{\boldsymbol{\theta\theta}}]_{7,1:7}=0\))  and neglecting $f_c$ covariance/correlation (i.e., by setting \([{\Sigma}_{\boldsymbol{\theta\theta}}]_{1:6,7}=[{\Sigma}_{\boldsymbol{\theta\theta}}]_{7,1:6}=0\) but keeping $\sigma^2_{\theta_7} \neq 0$ ). 

\begin{figure}[!htb]
\centering
   \includegraphics[width=0.48\textwidth]{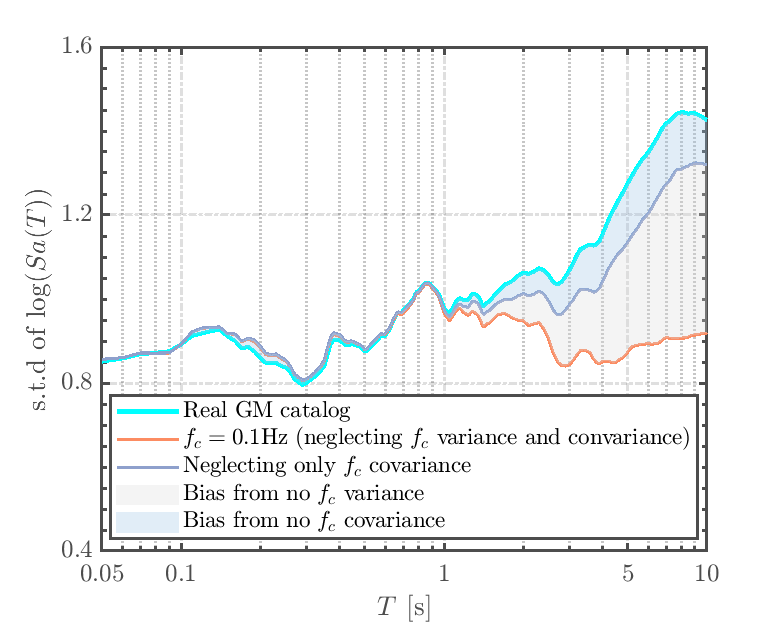}
   \caption{Effect of neglecting either the $f_c$ variance or the $f_c$ covariance on spectral variance.}
   \label{fig_7} 
\end{figure}

\begin{figure}[!htb]
   \includegraphics[width=0.95\textwidth]{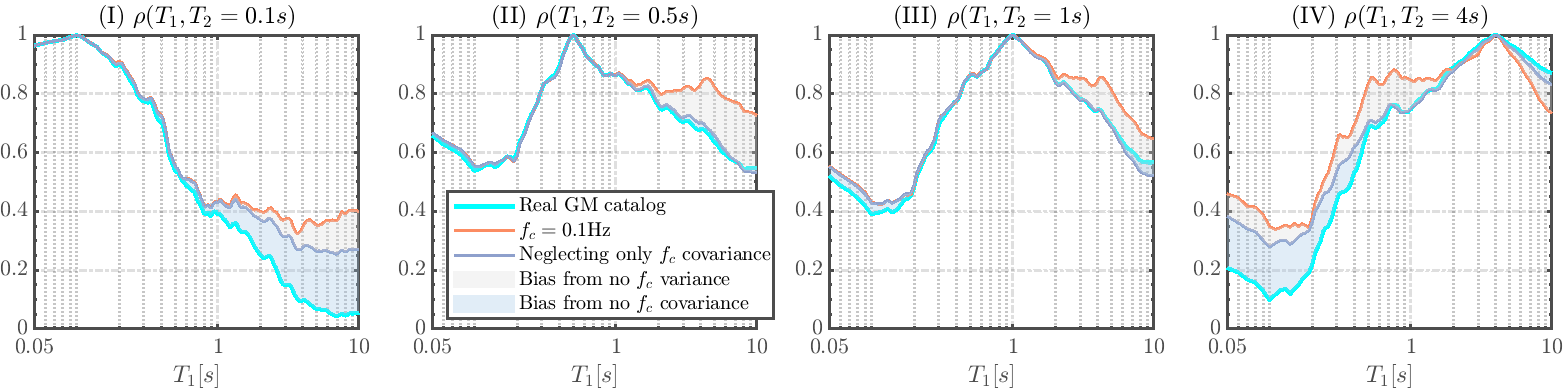}
\caption{Effect of neglecting either the $f_c$ variance or the $f_c$ covariance on spectral correlation.}
\label{fig_8} 
\end{figure}

Next,  we investigate the impact of the $f_c$  variability in the regression model by considering two scenarios: (1) neglecting both the $f_c$ variance and covariance by treating $f_c$ as a constant, e.g., $f_c=0.1$Hz, and (2) neglecting only the $f_c$ covariance. These two modifications change the spectral variance in Eq.\eqref{eq_sec3_2} and the spectral covariance in Eq.\eqref{eq_sec3_3}. The comparisons before and after the modifications are shown in Figure \ref{fig_7} and \ref{fig_8}.  Figure \ref{fig_7} suggests that neglecting either the $f_c$ variance or covariance leads to an underestimation of the spectral variance in the long-period range. The shaded areas in Figure \ref{fig_7} illustrate that the $f_c$ variance contributes more significantly than the $f_c$ covariance. Specifically, the $f_c$ variance can contribute up to a maximum of 28\%  s.t.d of $\log(Sa(T))$ to the total at $T=10$s, while the $f_c$ covariance contributes up to a maximum of 11\% at around $T=5.4$s. Figure \ref{fig_8} demonstrates  the impact of the $f_c$  variability on the spectral correlation, i.e.,  $\rho(T_1,T_2) = \operatorname{Cov}(Y(T_1),Y(T_2))/ \allowbreak \sqrt{\operatorname{Var}(Y(T_1))\operatorname{Var}(Y(T_2))}$.  It is evident that neglecting either the $f_c$ variance or covariance tends to overestimate the spectral correlation,  especially between the short- and long-period ranges. This outcome aligns with the finding in Figure \ref{fig2}, where improper $f_c$ values in the stochastic GM model result in overestimated correlations. The detailed analysis in Figure \ref{fig_8} reveals that the biased spectral correlation could be more influenced by the $f_c$ covariance than the $f_c$ variance.

In summary, the proposed linear regression accounts for most of the spectral variance and spectral covariance. Specifically, it shown that a fixed value of  \(f_c\) results in a distorted linear-response spectrum for long periods. Moreover, neglecting the covariance between \(f_c\) and the other six GM parameters severely underestimates spectral variance and significantly overestimates spectral correlation.

\section{Conclusions}
\noindent
This study emphasizes the importance of fitting the corner frequency (\(f_c\)) parameter in site-based stochastic GM models and proposes a method to optimize its fitting. Compared to previously used \(f_c\) estimation methods, GM models using the optimized $f_c$ show markedly enhanced accuracy in capturing the linear-response spectrum statistics of recorded GMs. A sensitivity analysis of the linear-response spectrum is conducted via linear regression. The regression incorporates seven GM parameters including $f_c$ as input variables. The analysis reveals that the value and variance of \(f_c\) and its correlation with the other six parameters contribute significantly to the linear-response spectral ordinates, variability, and correlation.

% The key findings and expectations in this paper are summarized as follows:
% \begin{itemize}
%     \item GM Model formulation in the time or frequency domain is not important
%     \item A linear structure found in the GM time series data
%     \item $f_c$ can be used as an Intensity measure to guide GM selection
%     \item The linear structure can explain why there is no remarkable difference between using Gaussian or vine-copula when concerning the spectrum variances and spectrum correlation
% \end{itemize}

\appendix

\section{Statistics of extracted parameters from the 71-GM catalog}
\label{AppendixA}
\begin{figure}[ht]
\centering
   \includegraphics[width=1\linewidth]{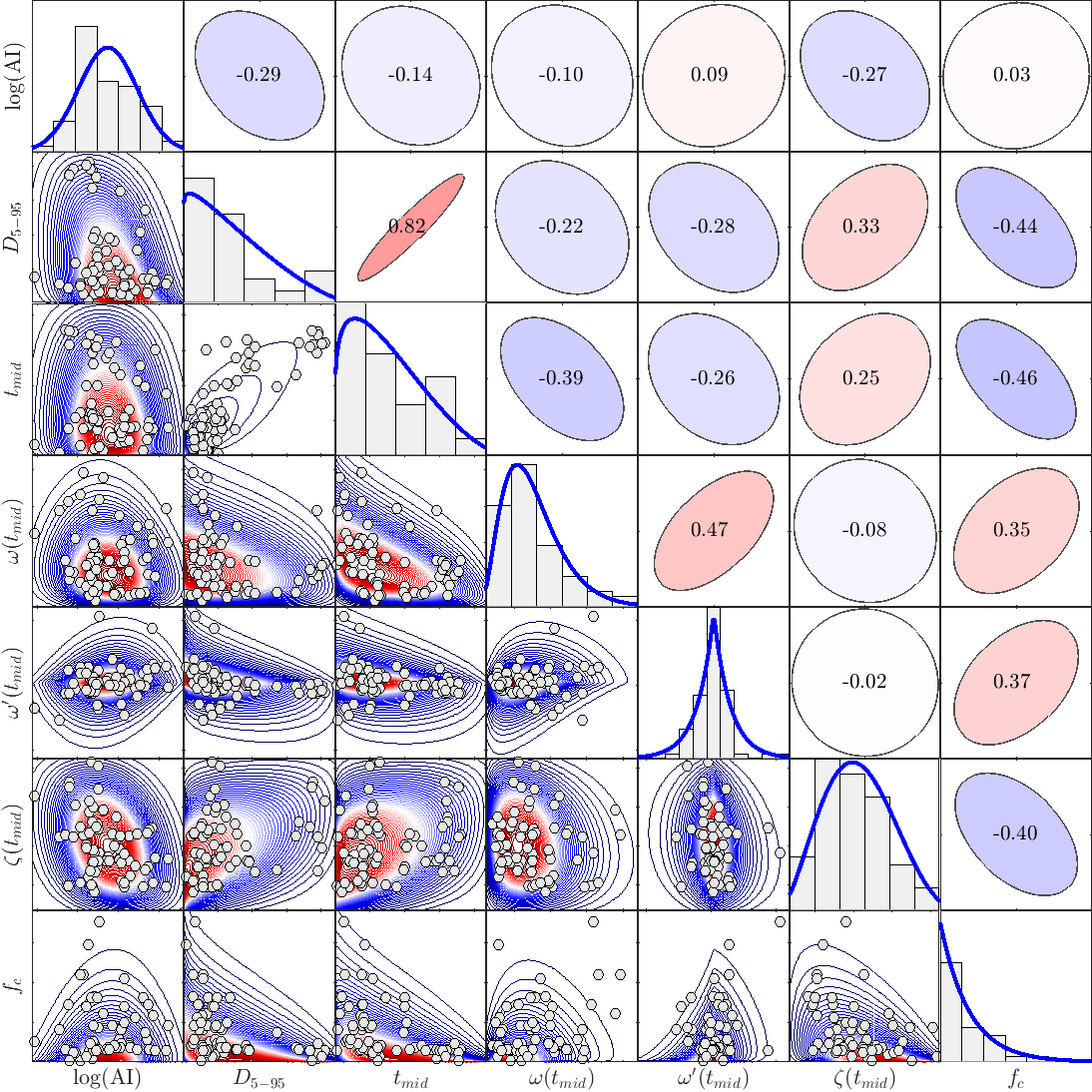}
   \caption{Statistics of seven GM parameters extracted from the 71-GM catalog.}
   \label{Appendix_A_fig} 
\end{figure}

\noindent
Figure \ref{Appendix_A_fig} presents the statistical analysis of the parameters extracted from the 71-GM catalog \cite{rezaeian_simulation_2012,broccardo2019preliminary}. In the diagonal positions, histograms of individual parameters are displayed alongside their fitted probability density functions. These marginal models for the GM parameters except for $f_c$ use the same distribution types as in the Rezaeian and Der Kiureghian model \cite{rezaeian_simulation_2012}; $f_c$ is fitted by an exponential distribution. The upper triangle exhibits Pearson's correlation coefficients for each pair of parameters. The lower triangle illustrates the bi-variate distribution incorporating Gaussian dependency, including their observed data points. 

\section{Percentages of spectral covariance decomposed by the regression model}
\label{AppendixB}
\begin{figure}[ht]
\centering
   \includegraphics[width=1\linewidth]{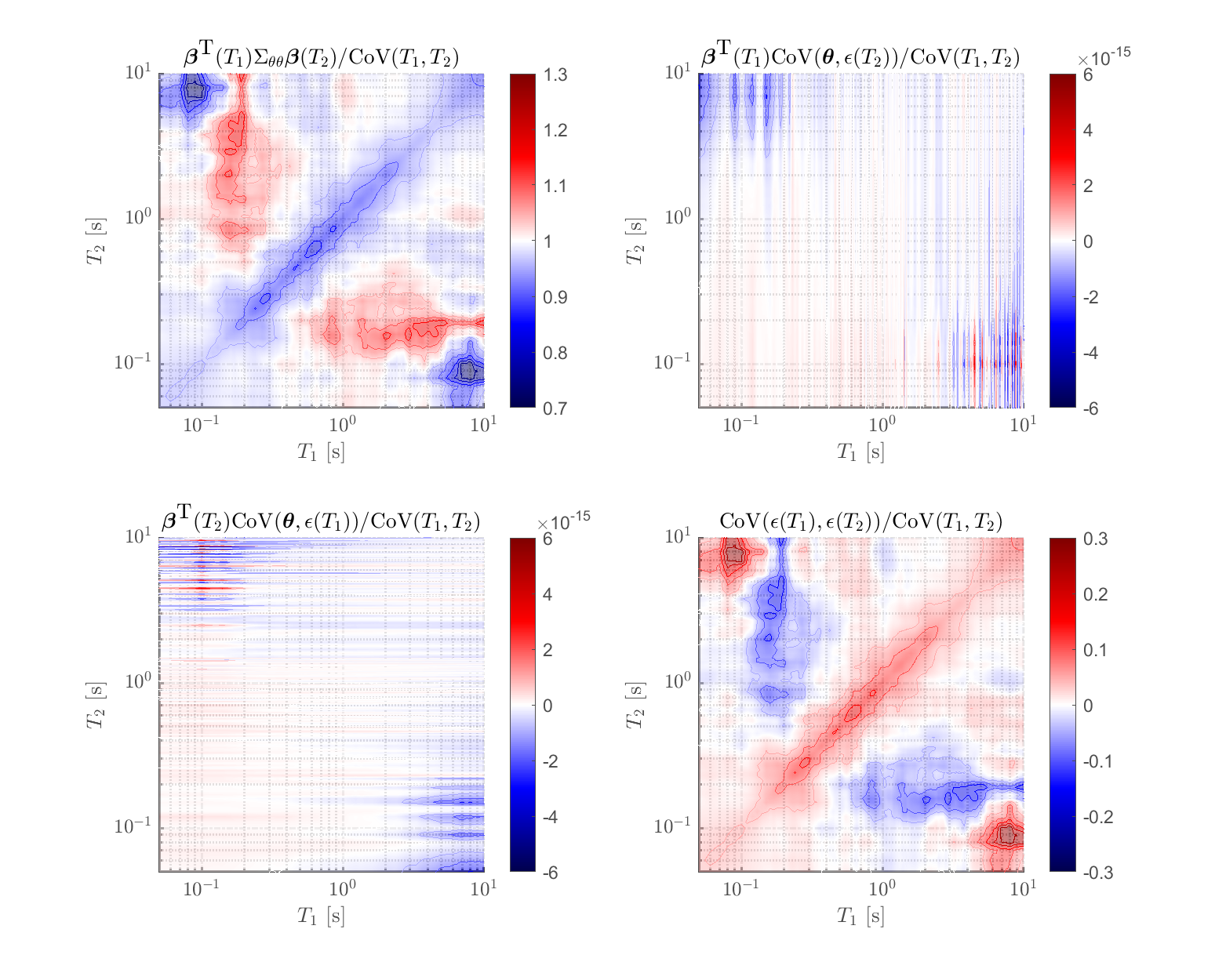}
   \caption{Percentage contributions of the four terms in Eq. \eqref{eq_sec3_3} to the total spectral covariance.}
      \label{Appendix_B_fig1} 
\end{figure}

\noindent
Figure \ref{Appendix_B_fig1} shows the percentages of the total spectral covariance explained by the four individual terms in Eq.\eqref{eq_sec3_3}.  The figure indicates that the linear-regression covariance, $\boldsymbol{\beta}^{\mathrm{T}}(T_1)\Sigma_{\boldsymbol{\theta\theta}}\boldsymbol{\beta}(T_2)$, accounts for $70\% \sim 130\%$ of the total covariance.  Contributions from the second and third terms, i.e., $\boldsymbol{\beta}^{\mathrm{T}}(T_1)\operatorname{Cov}(\boldsymbol{\theta},\epsilon(T_2))$  and $\boldsymbol{\beta}^{\mathrm{T}}(T_2)\operatorname{Cov}(\boldsymbol{\theta},\epsilon(T_1))$, are negligible. The percentages of the last term, i.e., covariance between regression errors at distinct periods, compensate for the linear-regression covariance almost to 100\%.

% \section{Additional appendix}
% \underline{Analysis to the Figure \ref{Appendix_B_fig2}}\\
% The new metric is defined as:
% \begin{equation}
%     {{\bar{E}}_{F}}({{f}_{0}})=\frac{\int_{0}^{{{f}_{0}}}{\ddot{a}_{g}^{2}}(f)df}{\int_{0}^{{{f}_{\max }}}{\ddot{a}_{g}^{2}}(f)df}\times 100,
% \end{equation}
% where $\ddot{a}_{g}(f)$ is the Fourier amplitude spectrum of a recorded GM. In Figure \ref{Appendix_B_fig2} (a), it shows the Spearman's correlation between $f_c$ and $\log({{\bar{E}}_{F}}({{f}_{0}})$. The result indicates when $f_0=0.2$Hz the new metric is most negatively correlated to $f_c$. In Figure \ref{Appendix_B_fig2} (b), it shows the histogram of $\log({{\bar{E}}_{F}}({{f}_{0}=0.2Hz}))$. Figure \ref{Appendix_B_fig2} (c) provides the scatter plot of $(f_c,\log({{\bar{E}}_{F}}({{f}_{0}=0.2Hz}))$ and conducts a linear regression fitting.
% Note that the results in Figure \ref{Appendix_B_fig2} are computed from the 756-GM strike-slip Catalog. The 71-GM catalog provides similar results.
% \begin{figure}[ht]
% \centering
%    \includegraphics[width=1\linewidth]{fig_apdix_B.png}
%    \caption{Correlation between $f_c$ and the new metric $\bar{E}_{F}({{f}_{0}})$}
%       \label{Appendix_B_fig2} 
% \end{figure}

\bibliographystyle{plain} % apalike
\bibliography{MyRef}

\end{document}